# COMMUTATIVE DIAGRAMS OF Sp(2) SYMMETRY

J. L. Vazquez-Bello[♦]

Physics Department
University of Puerto Rico
Bayamon Campus
Bayamon, PR-00959

**Abstract:** *It is well known that BRST symmetry plays a fundamental role in constructing quantum gauge theories. Yet, at the "classical" level, it constitutes the modern language to study constrained systems. First, this letter reviews the Sp(2) covariant quantisation of gauge theories, so that the geometrical interpretation of gauge theories in terms of quasi principal fibre bundles $Q(M_S, G_S)$ is the main scenario. It is then described the Sp(2) algebra for ordinary Yang-Mills theory. Some basic set theory and topology terms are reviewed to proceed then to use theorems on left inverse and right inverse of functions together with groups $(M_S, G_S)$ to have triangular composition of functions on manifolds. Henceforth, it is so the purpose of this letter to present topological structures to gauge theories, in particular to construct commutative diagrams for cocycle conditions of the Sp(2) symmetry.*

## 1. Gauge theories in terms of quasi-principal fibre bundles.

Gauge theories have a nice geometrical interpretation in terms of connections on a principal fibre bundle (*pfb*) $P(M, G)$, where $M$ is the base space-time manifold and $G$ is the gauge group [1,2,3,4]. However, it is known that quantisation of gauge theories requires the introduction of fields $(c^n{}_m, \boldsymbol{p}^n{}_m)$. It would be then desirable to have a formalism where those extra fields fit into some representation of a larger group where all the fields are components of a superfield. This is a step in the direction of recovering a geometrical interpretation of quantum gauge theories. The main ingredients in the construction of geometrical quasi-principal fibre bundles (*qpfb*) are a space-time base manifold $M$, a gauge group $G$, an extended superspace manifold $M_S$ which is obtained by adding two extra Grassmann variables $\theta^a$ $(a = 1,2)$ to $M$, in the case of *Sp(2)* symmetry,

---

[♦] e-mail: jlbellox@netscape.net
    j_vazquez@cutb.upr.clu.edu

and a supergroup $G_S$. The construction is performed basically in three steps [1,2,3]. It starts with a pfb $P(M,G)$ and extend the gauge group $G$ to a supergroup $G_S$. The composition of $G$ with a Grassmann algebra $B$ prolongs $P(M, G)$ to a pfb $P'(M, G_S)$. The most general supergroup $G_S$ can be represented in matrix form. In particular, *OSp(N/M)* groups are represented by block matrices of the form

1.1
$$\begin{pmatrix} A & E \\ C & D \end{pmatrix}$$

where *A, D* are $(N \times N)$ and $(M \times M)$ matrices whose elements are taken from the even part of the Grassmann algebra *B* constructed over a complex vector space *W*, whilst *C, E* are $(N \times M)$ and $(M \times N)$ rectangular matrices whose elements belong to the odd part of *B*. Next, it is enlarged the base space manifold *M* to a superspace $M_S$ in $P'(M, G_S)$ by adding Grassmann variables. At this stage, a pfb $P''(M_S, G_S)$ is obtained. Finally, the pfb $P''(M_S, G_S)$ is transform into a quasi-principal fibre bundle $Q(M_S, G_S)$. For instance, given a one-form valued function $\mathbf{a}(x) = A_m \, dx^m$ on *M* this induces a connection $\mathbf{w}$ on the pfb $P(M, G)$. Then a one-form valued function $\mathbf{a}'$ on $M_S$ is found by

1.2
$$\alpha'(x, \theta a) = g^{-1}(A_m \, dx^m)g + g^{-1}dg$$

where $g = g(x^m, \theta a)$, $(a = 1,2,\ldots)$ which induces a connection $\omega'$ on the qpfb $Q(M_S, G_S)$ [1,2,3,4].

## 2. The Sp(2) BRST Algebra of Yang-Mills Theory.

It has been realized for some time [5,6,7,8] that a geometrical construction can be useful for the discussion of BRST and anti-BRST symmetry[•]. The idea is to use an enlarged space with coordinates $Z^M = (x^m, q^a)$, where $(a = 1,2)$ and $q^a$ is an anti-commuting scalar coordinate and the BRST generators $s^a$ are realized as differential operators on superspace, $s^a = \{\partial/\partial q_a\}$, so that $s^a s^b + s^b s^a = 0$ holds automatically.

---

[•] {It is usually defined a bosonic operator $\sigma = (1/2)\varepsilon_{ab} s^a s^b$ where $\varepsilon_{ab}$ is the symplectic invariant form of *Sp(2)*, so that $\varepsilon_{ab} = -\varepsilon_{ba}$, $\varepsilon^{ab}\varepsilon_{bc} = \delta_{ac}$ and $\varepsilon_{12} = 1$. The generator $\sigma$ is invariant under *Sp(2)* and satisfies $s^a\sigma = 0$. The *Sp(2)* generators $\sigma^i$ (i = ±, 0) and the fermionic charges $s^a$ together form an algebra which is a contraction of *OSp(1, 1/2)* and denoted as *ISp(2)* [9,10]}.

For example, in Yang-Mills theory the gauge potential field $A^i_m$ and the Faddeev-Popov ghost field $(c^a)^i$ (where $i$ is an adjoint group index) can be combined into a super-gauge field $A^i_M(Z)$ whose lowest order components are

$$2.1 \quad [A_M^{\ i}(Z)]_{q=0} = (A_m^{\ i}, A^i_{\ q})_{q=0} = (A_m^{\ i}, (c^a)^i).$$

So that the standard Yang-Mills BRST transformations arise from imposing the constraints $F^i_{\ mq} = 0$, $F^i_{\ qq} = 0$, on the superfield strength $F^i_{MN}$ [5,6]. This gives an elegant geometrical description of BRST and anti-BRST symmetry.

Let us review the construction of gauge theories in the superspace with coordinates $Z^M = (x^m, \theta)$, which gives a geometric formulation of *Sp(2)* BRST symmetry. First, we consider matter fields $\Phi^i(x, \theta)$ and a gauge potential $A^i_{\ M}(x, \theta)$. These can be used to define a covariant derivative

$$2.2 \quad D_M \Phi^i = \partial_M \Phi^i - (T^k)^i_{\ j} A^k_{\ M} \Phi^j$$

and the field strength

$$2.3 \quad F^i_{\ MN} = \partial_M A^i_{\ N} - (-1)^{MN} \partial_N A^{Mi} + f_{j\ jk} A^j_{\ M} A^k_{\ N}$$

where $(-1)^{MN}$ is 1 unless both $M$ and $N$ are indices referring to anti-commuting coordinates, in which case it is $(-1)$. The gauge potential $A_M$ contains more component fields than the physical gauge and ghost fields and so, as in supersymmetric theories, constraints should be imposed on the field strength F[*]. Appropriate constraints are [5,6]

$$2.4 \quad F_{qq} = 0, \qquad F_{mq} = 0,$$

So that, the BRST and anti-BRST transformations corresponds to translations in the $\theta_a$ direction with $\partial/\partial \theta_a$ realized as differential operators on the enlarged space manifold $M_S$.

---

[*] {In Ref [6], it was obtained explicitly a geometrical formulation of BRST and anti-BRST symmetries and given the field content of $A^i_{\ M}(x, \theta)$ in terms of $\Lambda$, $\Upsilon$, $\Omega$ and $\omega$ fields. The components in the expansion can also be read as conditions on the mapping of some coordinates $\phi^i$ of the fibres over $U_i$ (covering set of $M_G$) and expressed as cocycle conditions.}

## 3. Basic Set Topology and Useful Theorems.

The part of geometry labeled topology deals with objects and properties among them. So then, let us first examine a number of those mathematical quantities like number systems, sets, universal covering spaces, spheres and projective spaces, to proceed and to borrow ingredients towards topological structures of gauge theories. We can then write commutative diagrams for cocycle conditions of Sp(2) symmetry. The above contrast with calculus where we usually strive mainly to acquire a competence in solving specific problems, or applications of some theorems to particular cases. It is generally understood that a theorem is often regarded as a *recipe* for applying formulas to special cases. With a concise set of definitions we shall be able to examine topological structures in general gauge models rather than just few cases.

*Definition.* A relation $f$ from A to B is called a function provided that for every $a \in A$ there is one and only one $b \in B$ such that $(a,b) \in F$, [10-16].

$$f: A \to B \qquad \text{or} \qquad A \xrightarrow{f} B$$

*Definition.* A function $f: A \to B$ is onto B ( or surjective ) if and only if $f[A] = B$; i.e. for every $b \in B$, there one $a \in A$ such that $b = f(a)$.

*Definition.* A function $f: A \to B$ is one to one (or injective) provided that $f(a) \neq f(b)$ whenever $a \neq b$.

*Proposition.* If $f$ and $g$ are functions, then $f \circ g$ is a function and for every $x$ in the domain of $f \circ g$, $(f \circ g)(x) = f(g(x))$.

*Proposition A.* If $f: A \to B$. Then the following statements are equivalent:

a) $f$ is one to one,
b) $f^{-1} \circ f = 1_A$
c) if $A \neq \emptyset$, then there exists some $g: B \to A$ such that $g \circ f = 1_A$
d) For every set $C$ and all pairs of functions $h: C \to A$ and $k: C \to A$ such that $f \circ h = f \circ k$ it follows that $h = k$ (i.e. $f$ is left cancelable with respect to composition).
e) $f^{-1}$ is a function from $f[A]$ to A.

*Proposition B.* If $f: A \to B$. Then the following statements are equivalent:

a) $f$ is onto $B$,
b) $f \circ f^{-1} = 1_B$
c) There exists some $g: B \to A$ such that $f \circ g = 1_B$
d) For every set $C$ and all pairs of functions $h: B \to C$ and $k: B \to C$ such that $h \circ f = k \circ f$, it follows that $h = k$ (i.e. $f$ is right cancelable with respect to composition).

*Definition.* A topology on a set X is a set $\tau$ of subsets of X :
a) $\emptyset \in \tau$ and $X \in \tau$,
b) if $U, V \in \tau$ then $U \cap V \in \tau$, and
c) if $\mathcal{U} \subseteq \tau$, then $\cup \mathcal{U} \in \tau$.

A topological space is a pair $(X,\tau)$, where $\tau$ is a topology on X if $(X,\tau)$ is a space, the elements of $\tau$ are called $\tau$ open sets, or just open sets, and the elements of X are called points of the space.

*Definition.* A subspace X of a topological space Y is a retract of Y if there is a continuous map (function) $r: Y \to X$ with $r(x) = x$ for all $x \in X$; such map $r$ is called a retraction.

Let us recall that in group theory, a group H contained in a group G is a subgroup of G if and only if the inclusion $i: H \to G$ is a homomorphism (this says that the group operations in H and G coincides).

In topology, a topological space X contained in a topological space Y is a subspace of Y if a subset V of X is open in X if and only if $V = X \cap U$ for some open subset U of Y. This guarantees that the inclusion $i: H \to G$ is continuous.

Now, let us suppose that there is a retraction $r: D^{n+1} \to S^n$, so there would be a "commutative diagram" of topological spaces and continuous maps

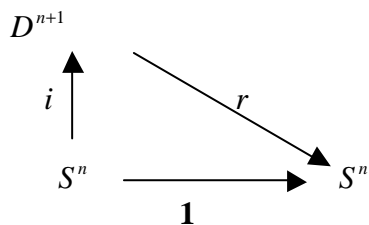

where commutative means that $r \circ I = 1$, the identity function on $S^n$. So let us recall proposition A and B on set theory [10-16].

*Definition.* A continuous surjection $f : X \to Y$ is an identification if a subset $U$ of Y is open if and only if $f^{-1}(U)$ is open in X.

*Theorem.* Let $f : X \to Y$ be a continuous surjection. Then $f$ is an identification if and only if for all spaces Z and all functions $g: Y \to Z$, one has $g$ continuous if and only if $gf$ is continuous

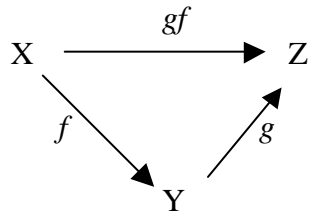

*Definition.* Let $f: X \to Y$ be a function and let $y \hat{I} Y$. Then $f^{-1}(y)$ is called the fiber over $y$.

*Corollary.* Let $f: X \to Y$ be an identification and, for some space Z, let $h: X \to Z$ be a continuous function that is constant on each fiber of $f$. Then $h f^{-1}: Y \to Z$ is continuous.

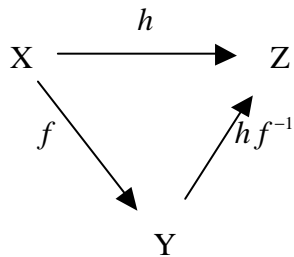

Moreover, $h f^{-1}$ is an open map (or closed map) if and only if $h(U)$ is open ( or closed ) in Z whenever $U$ is an open ( or closed) set in X of the form $U = f^{-1} f(U)$

    Let A, B be sets, such that $R \subset (A \times B)$ is understood to be a relation in A to B. Where properties like reflexive $(a,a) \in R$, symmetric $((a,b),(b,a)) \in R$, and transitive $(a,c) = ((a,b),(b,c)) \in R$, provides property of equivalence in R. For instance, the angle zero number is the same as the angle number $2\pi$, so that (x R y) iff (x,y) are two measurements of the same angle. Hence, equivalence classes (infinite) are expressed by $[0] = \{2k\pi : k \in Z\}$.

A function $f$ is then a especial kind of relation R. Let A, B be finite sets, such that A = Dom(f) and B = Rng(f). Then $f: A \to B$, so that $a \mapsto f(a)$. Henceforth, domain

Dom (R) = A and range Rng (R) = B are a relation in A to B, so that Im (R) is image of element *a* under relation R.

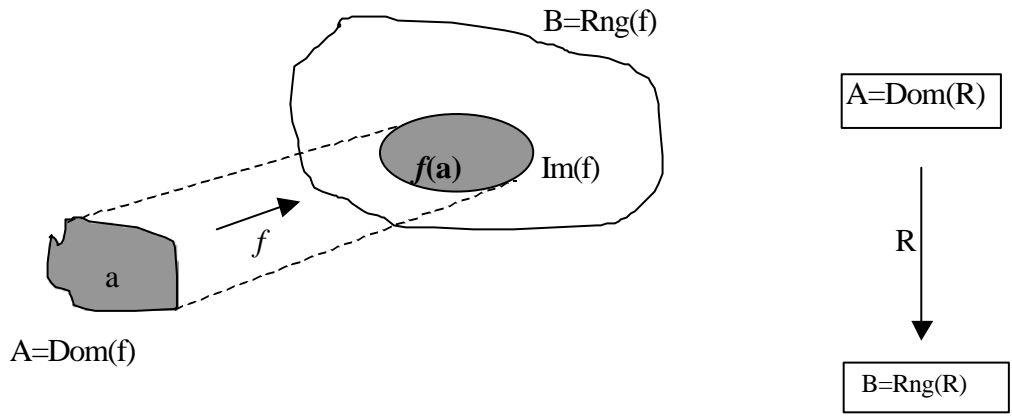

A large number of concepts in physics can be interpreted in terms of geometry of fibre bundles. For instance, Yang-Mills theory is a theory of connections on principle fibre bundles with gauge group G as a fibre [5,6].

**4. Commutative diagrams.**

Let us introduce some theorems of composition of functions and basic operations, based on previous propositions [10-16].

*Theorem A*. A function $f: A \to B$, where $A \neq \emptyset$, is 1-1 iff $f$ has left-inverse.

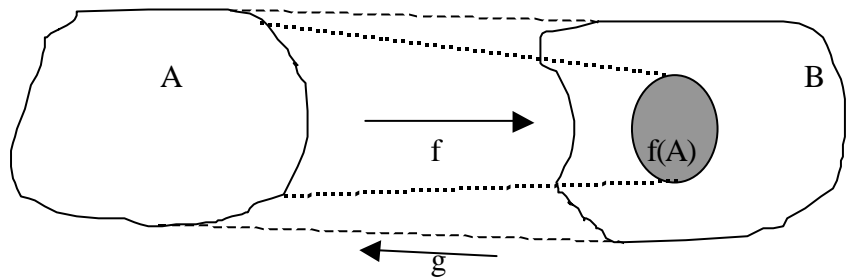

*Theorem B.* A function  $f: A \rightarrow B$ has a right-inverse iff Im($f$) = B and $f$ is *denoted* onto or surjective.

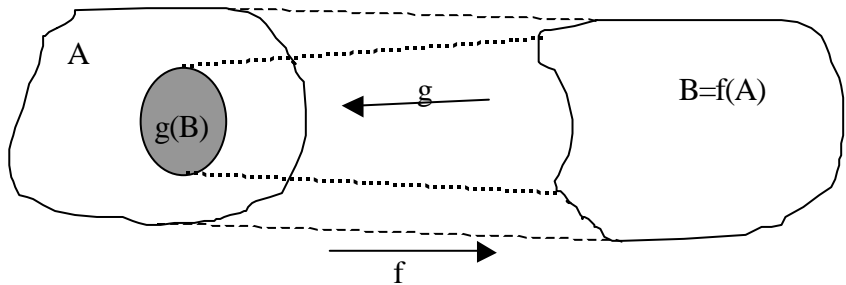

*Proposition.* Let A, B, C, D finite sets, and let $f, g, h$ well behaved $C^\infty$ functions. Then, the composition of functions is an associative binary operation.

$$A \xrightarrow{f} B \xrightarrow{g} C \xrightarrow{h} D$$

So that,

$$( h \circ g ) \circ f \; = \; h \circ ( g \circ f ).$$

Graphically,

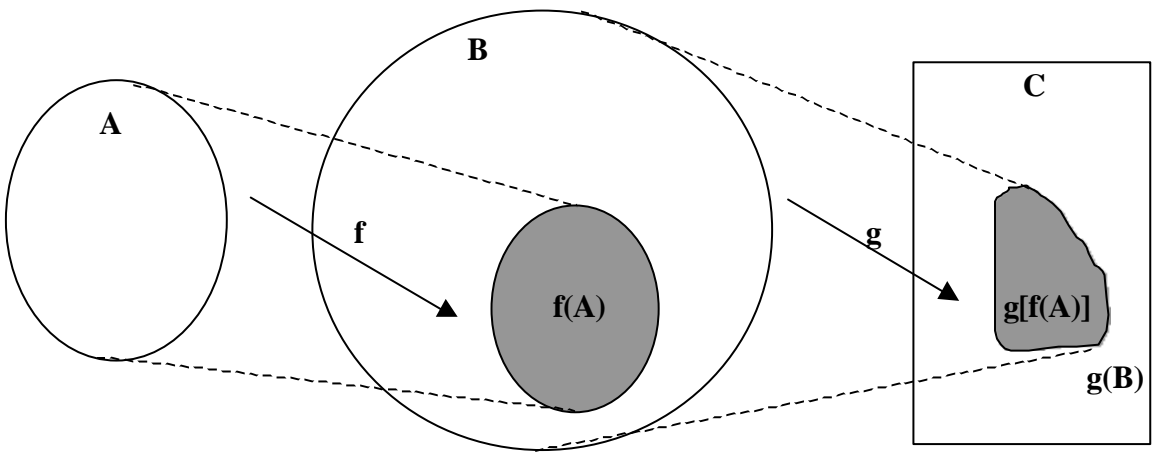

Let us define the following left-inverse commutative diagram to be the triangular composition of functions that satisfy theorem A. For left commutative diagram we have A, B sets and $f, g \in C^\infty$ functions, with unity. So that, $g$ composition $f$ gives the unique left-inverse unity $I_A \in C^\infty$ function, ($g \circ f = I_A$),

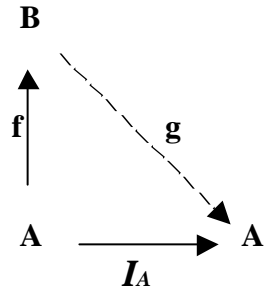

Let us define now the following right-inverse commutative diagram to be the triangular composition of functions that satisfy theorem B. For right-inverse commutative diagram we have A, B sets and $f, g \in C^\infty$ functions, with unity. So that, $f$ composition $g$ gives the unique right-inverse unity $I_A \in C^\infty$ function, ($f \circ g = I_B$),

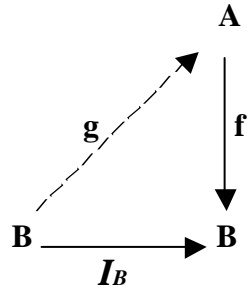

## 5. Cocycle conditions for Sp(2) symmetry.

We saw in previous sections that gauge theories have a nice geometrical interpretation in terms of connections on a principal fibre bundle (*pfb*) *P(M, G)*, where *M* is the base space-time manifold and *G* is the gauge group [17]. There components in the expansion for gauge field $A^i_M (Z)$ can be read as conditions on the mapping of some coordinates $\phi_i$ of the fibre bundle over $U_i$ (covering set of $M_G$) and expressed as cocycle conditions. Here, we used left-inverse commutative diagram to be the triangular composition of functions that satisfy theorem *A*, and right-inverse commutative diagram to be the triangular composition of functions that satisfy also theorem B.

Let us recall that many concepts in physics can be interpreted in terms of geometry of fibre bundles. For instance, Yang-Mills is a theory of connections on principle fibre bundles with a given gauge group G as fibre. So let M be a manifold that we shall call base manifold, let F be also a manifold called fibre. Then, a fibre bundle E over M with fibre F is a manifold with (M ⊗ F), lets say locally. If $\{U_i\}$ covers M then $(U_i \otimes F)$ topologically describes E, where $\phi_i$ are to be coordinates for patches $U_i$ on M that satisfy left inverse and right inverse commutative diagrams♦. Graphically, we have that

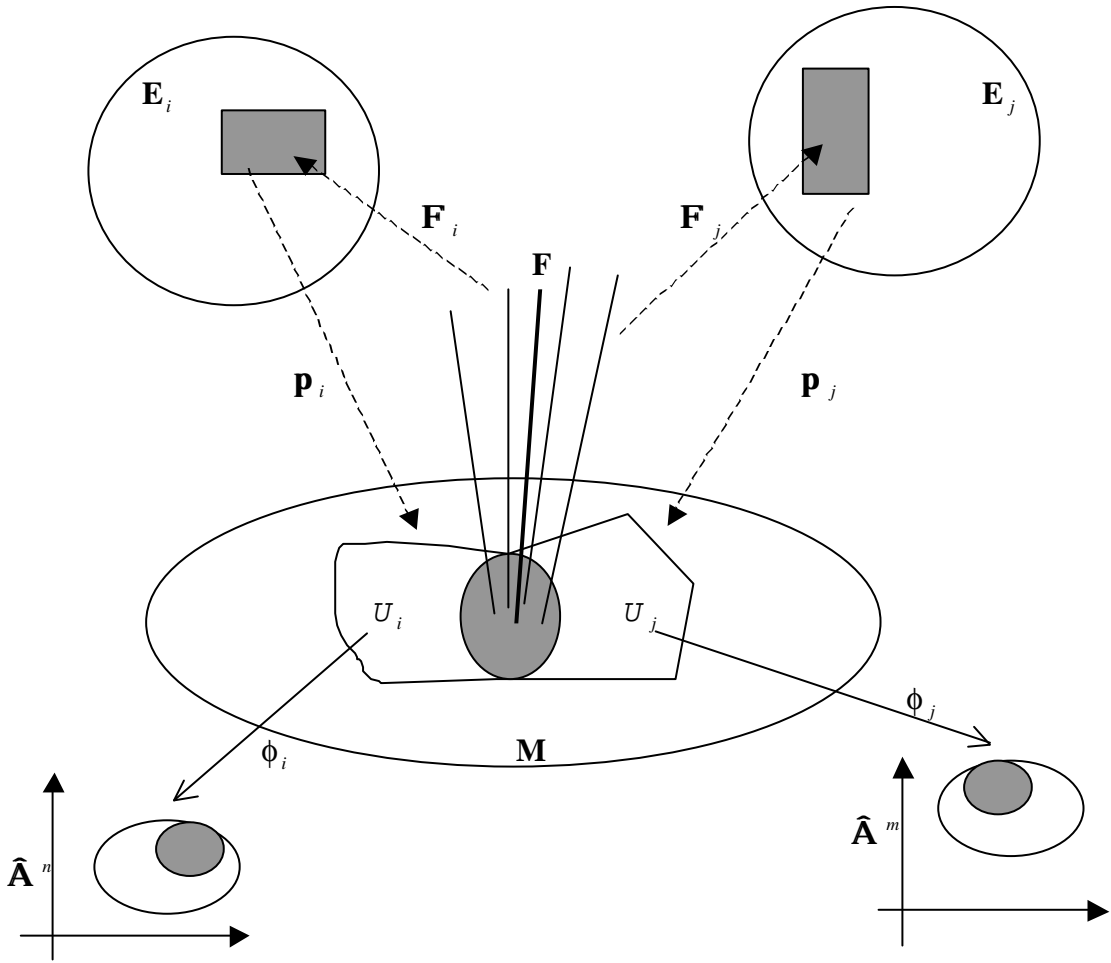

---

♦ Notation: We shall use compact indices on topological coordinates $\phi_i$ on $\{U_j\}$ so that $\phi_{ji} = \phi_j \phi_i^{-1}$, like a winding index.

Henceforth, $\Phi_{ij} = F_i \to F_j$ defines the mapping of the coordinates of the fibres over $U_i$, $U_j$ in intersection $U_i \cap U_j$ on base manifold M.

*Proposition.* Let $\Phi_i$ be a set of well-behaved functions on manifold (M $\otimes$ F) that describes the mapping of coordinates of the fibres over $U_i$ that covers M. Then cocycle conditions are compositions $\Phi_{ij}: F_i \to F_j$ in intersection $U_i \cap U_j$ on manifold M that satisfy

a) $\Phi_{ij}$ is the identity iff (i = j).
b) $\Phi_{ij} \Phi_{jk} = \Phi_{ik}$ is the composition.

This topological construction can be useful for discussion of a geometrical construction of BRST and anti-BRST symmetry. So that, if we identify symplectic invariant forms of Sp(2) as the mapping of the coordinates of the fibres manifold and fibre bundles E over base manifold M, either left invariant or right invariant commutative to the triangular composition of functions (coordinates). The cocycle conditions are the equivalents of the BRST transformations corresponding to translations in the θ direction on enlarged spaces ($x^m$, $q^a$), see section 2. Interestingly, the topological structure of cocycle conditions include the general case of mixed BRST and anti-BRST symmetry, when the global topology on the covers for base manifold M presents twisting of fibres on ($U_i \cap U_j$), and both symmetries are then coupled. There, the enlarged space will include ($x^m$, $q^a, \bar{q}^a$) space coordinates. In general, the more twisting on fibre manifold F, the more BRST's for BRST's symmetry, and possibly anti-BRST's for anti-BRST's symmetry too [6]. This process could continue indefinitely, giving and infinite sequence of BRST's for BRST's symmetries, if right inverse and left inverse commutative diagrams are Involved in the global topology with many twisting.

*Acknowledgments. It is a great pleasure to thank those authors on topology books for suggestions made on research and connections with physics. I wish to thank a number of non-specialists for reading and suggestions to this letter. I also place my gratitude to the physics department for their support. This piece of work is dedicated to my beloved wife and daughters whatever bundles they might have, yet.*

**References.**